\documentclass[twocolumn]{aastex631}
\usepackage[all]{hypcap}
\usepackage[utf8]{inputenc}
\usepackage{amsmath,ulem}
\newcommand{\out}{\rm out} 

\newcommand{\crit}{\rm crit}
\newcommand{\iom}{\mathcal{I}}
\newcommand{\ene}{\mathcal{E}}
\newcommand{\rout}{\langle r_{\out}\rangle}
\newcommand{\ie}{(\iom,\,\ene)}
\newcommand{\rzv}{t_{\rm ZLK}/t_{\rm VRR}}
\newcommand{\rze}{t_{\rm ZLK}/t_{\rm evap}}

\newcommand{\rmax}{\mathcal{R}_{\max}}

\newcommand{\qinmin}{q_{{\rm in,\,}\min}}

\begin{document}

\title{Dynamical evolution of stellar binaries in galactic centers}
\shorttitle{Stellar binaries orbiting MBHs}
\shortauthors{Dodici, Tremaine, \& Wu}

\author[0000-0002-3352-9272]{Mark Dodici}
\affiliation{Department of Astronomy \& Astrophysics, University of Toronto, Toronto, ON M5S 3H4, Canada}
\affiliation{Canadian Institute for Theoretical Astrophysics, Toronto, ON M5S 3H8, Canada}
\email{mark.dodici@astro.utoronto.ca}

\author[0000-0002-0278-7180]{Scott Tremaine}
\affiliation{Canadian Institute for Theoretical Astrophysics, Toronto, ON M5S 3H8, Canada}
\affiliation{Institute for Advanced Study, Princeton, NJ 08540, USA}

\author[0000-0003-0511-0893]{Yanqin Wu}
\affiliation{Department of Astronomy \& Astrophysics, University of Toronto, Toronto, ON M5S 3H4, Canada}

\date{\today}

\begin{abstract}
    Stellar binaries in galactic centers are relevant to several observable phenomena, including hypervelocity stars, X-ray binaries, and mergers of stars and compact objects; however, we know little about the properties of these binaries. Past works have suggested that a small fraction of them should contract to a few stellar radii or collide, due to the co-operation of stellar tides and the eccentricity oscillations induced by the strong tidal field of the central massive black hole.
    We revisit this model with several updates. 
    We first argue that when a binary's pericenter separation is driven down to a few stellar radii, diffusive excitation of stellar tides should quickly contract the orbit, saving the stars from collision. Instead, the stars should end up as a very tight binary. We then show that vector resonant relaxation and perturbations from passing stars --- effects not included in past models --- dramatically increase the prevalence of such encounters. 
    In numerical experiments, we find that 1 in 5 binaries around $1$ pc from Sgr A* should tidally contract in this way while still on the main sequence. This rate climbs to 3 in 5 around $0.01$ pc, inward of which it plateaus.
    We briefly discuss observable implications of these results, with particular attention to young stellar binaries in the Galactic Center.
\end{abstract}

\section{Introduction}

Stellar binaries orbiting massive black holes (MBHs) are involved in the production of several observable populations in our own Galactic Center, as well as a range of transient events in the centers of other 
galaxies.\footnote{For clarity, we hereafter define a ``galactic center'' as the region around an MBH containing an extended mass of order the MBH mass; in the Milky Way, this region's radius is $\approx 2$ pc.}
The former include hypervelocity stars \citep{Hills_1988,Yu_2003,Koposov_2020}, the S-star cluster \citep{Gould_2003,Ginsburg_2006,Generozov_2020}, G2-like objects \citep{Gillessen_2012,Prodan_2015,Stephan_2016}, and X-ray binaries \citep{Hailey_2018,Stephan_2019,Mori_2021}. The latter may include (kilo/super)novae \citep[e.g.,][]{Ginsburg_2007,Antonini_2010,Antonini_2011,Stephan_2019}, quasi-periodic eruptions \citep[e.g.,][]{Wang_2022,Linial_2023b,Lu_2023}, and gravitational wave signals from compact object binary mergers \citep[see, e.g.,][for overviews]{Tagawa_2020,ArcaSedda_2020,ArcaSedda_review_2023}.

Despite their roles in a range of astrophysical phenomena, we have scant information about these binaries.
Even in our own Galactic Center, we have only observed four spectroscopically confirmed stellar binaries within the central parsec (\citealp{Ott_1999,Pfuhl_2014,Gautam_2024}; see also \citealp{Chu_2023}).
Some studies have used the previously mentioned observables (e.g., the S-star cluster in \citealp{Generozov_2020} and hypervelocity stars in \citealp{Evans_2022a}) to constrain binary population properties; however, significant statistical and systematic uncertainties hinder such studies.

With observational constraints of galactic center stellar binaries in their infancy, we aim to refine theoretical expectations for this 
population.

In particular, we focus on one key process that shapes this binary population: \emph{tidal friction}. 
Briefly, when the members of a binary have a small separation at pericenter, they raise tides on each other which sap energy from their orbit. 
This process may reduce the binary's semimajor axis substantially \citep[e.g.,][and many others]{Zahn_1977,Hurley_2002,Ogilvie_2014}. We will focus on friction in the so-called ``diffusive tide'' regime, which is relevant at pericenter separations smaller than a few stellar radii \citep[e.g.,][]{Kochanek_1992,Mardling_1995a,Mardling_1995b,Mardling_2001,Ivanov_2004,Vick_2018,Wu_2018}.

Our galactic center binaries should commonly reach this regime. The strong tidal gravity here --- primarily from the MBH --- causes binaries to undergo extreme eccentricity oscillations. 
Except for slight deviations caused by the tidal potential of the cluster \citep{Hamilton_2019a,Hamilton_2019b}, these are the the classic von Zeipel-Lidov-Kozai (ZLK) cycles (\citealp{vonZiepel_1909,Lidov_1962,Kozai_1962}; see also \citealp{Naoz_2016,Tremaine_2023}). 
Past studies have found that such oscillations, coupled with tidal friction, should cause a small fraction of galactic center binaries to contract or to collide \citep{Antonini_2010,Antonini_2011,Antonini_2012,Prodan_2015,Stephan_2016,Bradnick_2017,Stephan_2019,Fragione_2019}.
For example, \citet{Stephan_2016,Stephan_2019} found that about $1$ in $10$ binaries in the inner 0.1 pc of the Galactic centre should tidally contract while on the main-sequence (MS), and that a comparable fraction should collide at very large eccentricities during these oscillations.

In this paper, we argue that the contracted fraction should be much larger, while the colliding fraction of MS binaries should be negligible. These differences arise naturally from two major updates to past models.

First, we emphasize that all binaries that would have collided in \citet{Stephan_2016,Stephan_2019} must first pass through the aforementioned diffusive tide regime. There, energy in the stars' fundamental modes grows with successive pericenter passages, rapidly reducing the orbital binding energy. As a result, the binary orbit contracts much faster than in the equilibrium-tide models used in \citet{Prodan_2015}, \citet{Stephan_2016,Stephan_2019}, and many other past 
works. We argue that this quick contraction should decouple the binaries from external perturbations before direct collisions can occur.

Second, we include two more external effects in our models --- namely,  gravitational perturbations by passing stars \citep[flybys; e.g.,][]{Collins_2008,Hopman_2009,Michaely_2020,Hamilton_Modak_2024} and vector resonant relaxation of the binary orbit about the MBH (VRR; e.g., \citealp{Rauch_1996,Kocsis_2011,Kocsis_2015}).
We show that these processes significantly enhance the fraction of binaries undergoing large eccentricity oscillations. This was previously discussed in the context of compact-object binaries in galactic centers by \citet{Hamers_2018} and \citet{Winter-Granic_2024}. 
Those works emphasize the relevance of flybys and VRR in the narrow regimes where they act on timescales comparable to the period of a ZLK cycle; for our problem, we find that these effects are important everywhere, and particularly in the much-broader regime where oscillations are faster. 

Combining these effects, we find that of order 1 in 2 stellar binaries in a galactic center should contract to near-contact separations while still on the MS.

In Section~\ref{sec:dynamics}, we argue that contraction should be more likely than collision during extreme eccentricity oscillations. In Section~\ref{sec:dynamics_2}, we discuss the interplay between ZLK oscillations, flybys, and VRR. We then numerically evolve a population of binaries in the Galactic 
Center (Section~\ref{sec:setup}) and report results in Section~\ref{sec:results}. In Section~\ref{sec:discussion}, we briefly discuss the observable implications of a significant population of near-contact binaries in galactic centers, and we conclude in Section~\ref{sec:conclusion}.

\begin{figure}
    \centering
    \includegraphics[width=\linewidth]{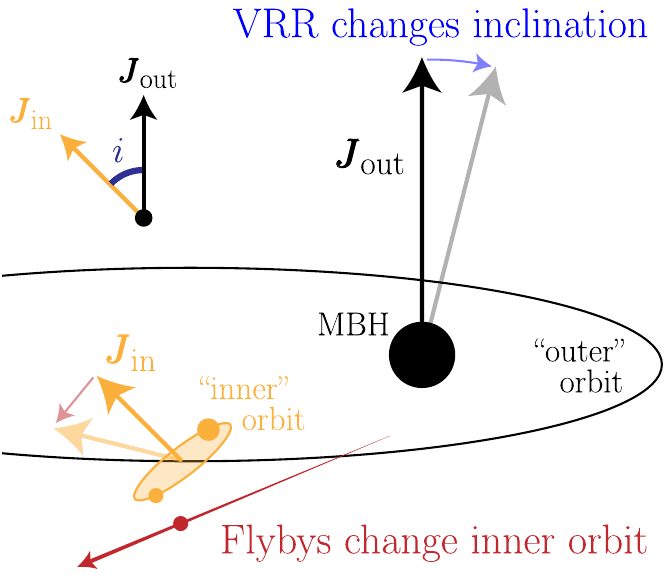}
    \caption{Stellar binaries in galactic centers form a hierarchical triple with an MBH, leading to oscillations of the inner-orbit eccentricity. For systems in a favorable region of parameter space (see Fig.~\ref{fig:ie_space}), these oscillations bring the inner orbit into the diffusive-tide regime, where it should rapidly contract (Section~\ref{sec:dynamics}). Flybys and VRR bring more systems into this favorable region (Section~\ref{sec:evolution}); flybys impulsively alter all properties of the inner orbit, while VRR continuously changes the mutual inclination $i$.}
    \label{fig:diagram}
\end{figure}

\section{DIFFUSIVE TIDAL FRICTION}\label{sec:dynamics}

We consider a binary of total mass $m_b$ orbiting an MBH of mass $m_\bullet = 4\times10^6\,M_\odot$, as depicted in Figure~\ref{fig:diagram}. We refer to the motion of the binary members about their barycenter as the ``inner'' orbit and the motion of their barycenter about the MBH as the ``outer'' orbit. 
For each orbit, we attach the subscripts ``in'' and ``out,'' respectively, to orbital properties including periods $P$, semimajor axes $a$, eccentricities $e$, and pericenter separations $q$. We define the inclination $i$ as the angle between the inner- and outer-orbit angular-momentum vectors ($\pmb{J}_{\rm in}$ and $\pmb{J}_{\out}$, respectively). We will refer often to the inner-orbit binding energy, $E_{\rm in} \equiv - Gm_b/2a_{\rm in}$.

\subsection{Oscillation and contraction}\label{sec:tidal_diff}

Binaries in galactic centers undergo oscillations in $e_{\rm in}$ driven by the strong tidal field. 
These roughly follow the classical ZLK oscillations (see introduction).
Over a single oscillation, an inner-orbit's pericenter separation evolves smoothly between maximum and minimum values, while its semimajor axis is conserved. 

We consider those systems reaching minimum pericenter separations smaller than a few stellar radii.
These orbits almost always satisfy $1-e_{\rm in} \ll 1$, so one can approximate that the stars only tidally interact during a pericenter passage.
We focus on the orbital evolution due to repeated excitation of the dynamical tides at every pericenter passage.

At each passage, one can abstract the tidal interaction as a ``kick'' of energy delivered to the stars' fundamental- or f-modes --- the oscillatory response of the star with zero radial nodes.
This energy comes from the orbit, so the binding energy changes by some $\Delta E_{\rm in}$; the inner-orbital period is then adjusted by $\Delta P_{\rm in} = (3/2)P_{\rm in}(\Delta E_{\rm in}/|E_{\rm in}|)$.

\begin{figure} 
    \centering
    \includegraphics[width=\linewidth]{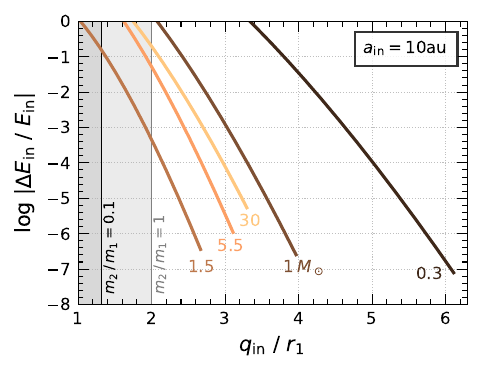}
    \caption{
    The orbital energy change after a single pericenter passage drops steeply with increasing pericenter separation (eq.~\ref{eq:delta_E}). Profiles are shown in the diffusive regime, where $|\Delta P_{\rm in}|\geq \omega_f^{-1}$. Each colored line is for a different primary mass $m_1$, labelled in $M_\odot$.
    The fractional change is $\propto a_{\rm in}$; here we show $a_{\rm in} = 10$ au. 
    Shaded regions show the sum of stellar radii for the listed mass ratios, assuming $r_2=r_1(m_2/m_1)^{1/2}$.}
    \label{fig:n_to_shrink}
\end{figure}

For a star initially not oscillating, one can calculate the energy change per passage as in \citet{Press_1977}; for an overview, see Appendix~\ref{sec:tidal_onset}. The fractional change is plotted in Figure~\ref{fig:n_to_shrink}.
It is a very steep function of the pericenter distance --- for instance, the fractional change can rise by an order of magnitude when the pericenter distance drops by $10$ percent.

Over many passages, these impulsive energy changes can add up diffusively when consecutive pericenter passages are not strictly periodic, but shift by at least of order one f-mode period.
That is, when $|\Delta P_{\rm in}| \gtrsim \omega_f^{-1}$ ---
with $\omega_f$ the f-mode frequency ---
the mode amplitude grows in a random walk, as the changes in period cause successive kicks to the mode to arrive at effectively random phases \citep{Ivanov_2004,Vick_2018,Wu_2018}. For our problem, we assume that $\Delta P_{\rm in}$ is caused purely by the loss of orbital energy to the tides, though there may be other drivers of period change.

\begin{figure}
    \centering
    \includegraphics[width=\linewidth]{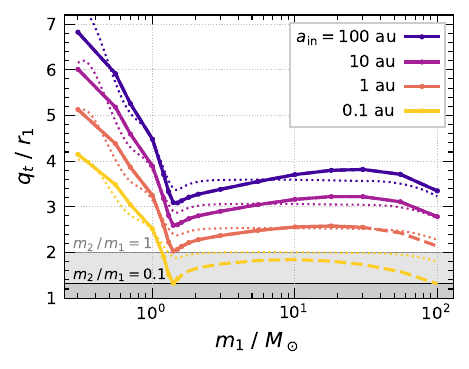}
    \caption{For typical semimajor axes, diffusive tides begin at wider $q_{\rm in}$ than collisions. This is not necessarily true at very small $a_{\rm in}$, though in this regime our tidal calculations assuming $1-e_{\rm in} \ll 1$ likely underestimate the fractional energy change per orbit. The dependence on $m_1$ arises solely from stellar structure. 
    Thin, dotted lines show the approximation~(\ref{eq:qt}). Lines are dashed in regions where $1-e_{\rm in} \not\ll 1$. Shaded regions show the sum of stellar radii for the listed mass ratios, as in Figure~\ref{fig:n_to_shrink}.}
    \label{fig:qt_num}
\end{figure}

We define $q_t$ as the pericenter separation at which this condition is first satisfied, i.e., where $|\Delta P_{\rm in}| = \omega_f^{-1}$. We calculate $q_t$ numerically for a range of primary masses $m_1$ and at several $a_{\rm in}$ (see Appendix~\ref{sec:tidal_onset}). The resultant profiles are shown in Figure~\ref{fig:qt_num}. These can be reasonably approximated as
\begin{equation}\label{eq:qt}
    q_t(m_1,a_{\rm in}) \approx 0.013 \;{\rm au}\left(\frac{m_1}{M_{\odot}}\right)^{0.55}\left[1 + \frac{1}{11}\ln\left(\frac{a_{\rm in}}{\rm au}\right)\right].
\end{equation}
This approximation should hold for any main-sequence star with a companion on a highly eccentric orbit (neglecting any tidal response in the companion). 

When a system is in this ``diffusive'' growth regime ($q_{\rm in} \leq q_t$), the mode energy increases roughly in proportion to the number of pericenter passages \citep[e.g.,][]{Mardling_1995a,Vick_2018,Wu_2018}. The expected semimajor axis after $N$ passages is then
\begin{equation}\label{eq:new_a}
    {\langle a_{{\rm in,}\,N}\rangle}\sim  a_{{\rm in,}\,0}\,{\left(1 + N\bigg|\frac{\Delta E_{\rm in}}{E_{{\rm in,}\,0}}\bigg|\right)}^{-1},
\end{equation}
where $0$ denotes a value prior to diffusive evolution.

This contraction, like other models of tidal friction, reduces $a_{\rm in}$ while roughly conserving the orbital angular momentum. This means $q_{\rm in}$ remains roughly constant, as long as the orbit is highly eccentric. 
The inner orbit contracts until $q_t (a_{\rm in})$ becomes smaller than $q_{\rm in}$ once again. At this point, the diffusive tide stalls. The succeeding tidal evolution is currently unclear \citep[see discussion in, e.g.,][]{Wu_2018}. We believe it is likely that the inner binary, now dynamically detached from the external disturbances, will continue to circularize over time.

When this tidal friction is coupled with ZLK oscillations, the result of a visit to the diffusive regime depends strongly on the minimum pericenter separation. For example, consider a binary with $m_1=1\,M_\odot$, $m_2 \leq m_1$, and $a_{\rm in} = 100$ au. The diffusive regime for this binary begins at $q_{\rm in} \approx 4.5\, r_1$ (see Fig. \ref{fig:qt_num}), where $r_1$ is the radius of $m_1$. If the ZLK oscillation is able to deliver the orbit to a pericenter separation of $4r_1$, the binary can contract to $a_{\rm in} = 10$ au before exiting the diffusive regime. In contrast, if the orbit reaches $q_{\rm in} = 2.5r_1$, it can contract all the way to $a_{\rm in} = 0.1$ au before the diffusive tide gives up. If the binary reaches $q_{\rm in} < r_1 + r_2$, the member stars will physically collide. While a simple parameter space argument implies that the majority of systems reaching $q_t$ will also reach $r_1+r_2$ (see also Section~\ref{sec:ZLK_loss}), we suggest in the following section that such collisions are unlikely.

\subsection{Fates in the diffusive regime}\label{sec:fates}

We give a simple argument that binaries driven into the diffusive regime by ZLK oscillations 
will circularize and contract, rather than being driven to direct collision. 

First, we define the range of inner orbits relevant to our problem. In the tidal field of an MBH, the widest stable binaries have semi-major 
axes\footnote{This approximation is equivalent to requiring a binary with $e_{\rm in} \rightarrow 1$ to have a separation smaller than its Hill radius at all times \citep[cf.][]{Grishin_2017,Vynatheya_2022}.}
\begin{align} \nonumber
    (a_{\rm in})_{\rm wide} &= \frac{1}{2}\, a_{\rm out}(1-e_{\rm out})\left(\frac{m_b}{3m_\bullet}\right)^{1/3}
    \\ \label{eq:a_wide}
    &\sim 57\,{\rm au}
    \left(\frac{a_{\out}}{0.1 \,{\rm pc}}\right)
    \left(\frac{m_b}{2\,M_\odot}\right)^{1/3},
\end{align} 
taking $e_{\rm out}=0$ for the scaling relation.

If the widest binaries can contract rather than collide, we expect all smaller ones to do so as well.
The widest binaries undergo the fastest eccentricity oscillations --- the ZLK cycle period is 
roughly\footnote{The precise cycle period is typically a factor of a few larger than equation~(\ref{eq:t_osc}), depending on the conserved values of a ZLK oscillation \citep[see, e.g.,][]{Antognini_2015,Basha_2025}. The period definition used here is precise for librating cycles in the limit of very small amplitudes
--- see Figure 1 of \citet{Antognini_2015}, noting that our definition is precise when the plotted value is $5/4$.} 
\citep[e.g.,][]{Eggleton_1998} 
\begin{equation}\label{eq:t_osc}
    t_{\rm ZLK} = \frac{2}{3\pi} \frac{P_{\rm out}^2}{P_{\rm in}}(1-e_{\rm out}^2)^{3/2}.
\end{equation}

How slow must an oscillation be before contraction is the expected outcome?
A binary spends a fraction $\sim (2q_{\rm in}/a_{\rm in})^{1/2}$ of each ZLK oscillation with pericenter separation smaller than a given $q_{\rm in}$ (\citealp{Anderson_2016}). This corresponds to $\sim(t_{\rm ZLK}/P_{\rm in})(2q_{\rm in}/a_{\rm in})^{1/2}$ pericenter passages.
In the diffusive regime, the inner orbit takes $\sim |E_{{\rm in},\,0}/\Delta E_{\rm in}|$ pericenter passages to contract by order unity (eq.~\ref{eq:new_a}).
As argued in \citet{Vick_2019}, diffusive tides decouple the inner orbit from ZLK oscillations at the pericenter separation where these numbers are comparable, i.e., where
\begin{equation} \label{eq:vick_lai_condition}
    q_{\rm in}\bigg|\frac{\Delta E_{\rm in}}{E_{{\rm in},\,0}}\bigg|^2 \sim \frac{a_{\rm in}}{2}\left(\frac{P_{\rm in}}{t_{\rm ZLK}}\right)^2.
\end{equation}

With this equality, we can finally determine the widest binary that will contract rather than collide. We rewrite equation~(\ref{eq:vick_lai_condition}) as a function of semimajor axis by evaluating the left hand side at $q_{\rm in} = r_1+r_2$. To do so, we parameterize the fractional one-kick energy change at this separation as $|\Delta E_{\rm in} / E_{{\rm in,}\,0}|_{r_1+r_2} = \delta E\times(a_{\rm in}/1\,{\rm au})$, where $\delta E$ depends on primary mass and binary mass ratio (see Fig.~\ref{fig:n_to_shrink}).
Then we find
\begin{align}
    \left(a_{\rm in}\right)_{\rm crit} &\sim 52\,{\rm au}
    \left(\frac{\delta E}{10^{-0.8}}\right)^{2/5}
    \left(\frac{a_{\out}}{0.1 \,{\rm pc}}\right)^{6/5}
    \nonumber \\ \label{eq:a_crit}
    &\quad\quad\quad\quad\quad
    \times 
    \left(\frac{m_b}{2\,M_\odot}\right)^{2/5}
    \left(\frac{r_1+r_2}{2\,R_\odot}\right)^{1/5}.
\end{align}
Any binary smaller than this will be expected to contract rather than collide. 

Comparing scaling laws~(\ref{eq:a_wide}) and (\ref{eq:a_crit}), we see that diffusive tides will save almost all stable binaries from collision. From their relative dependencies on $a_{\rm out}$, this statement is less true closer to the MBH. It will also be less true for binaries with more-massive primaries, as the appropriate $\delta E$ will be smaller (Fig.~\ref{fig:n_to_shrink}). That said, we also note that binaries with more unequal mass ratios have much wider $(a_{\rm in})_{\rm crit}$, as decreasing $r_2/r_1$ slightly yields a drastic increase in the fractional kick energy at $q_{\rm in}=r_1+r_2$ ($\delta E$; see Fig.~\ref{fig:n_to_shrink}).

Even if a binary is not saved by this argument, ZLK oscillations may be suppressed prior to collision through other precessional effects. Each pericenter passage induces a change to the argument of pericenter from rotational bulges, general relativity, and from the f-mode excitation itself. These so-called ``short-range forces'' may yield precession on timescales shorter than the ZLK timescale, in which case they may disrupt ZLK oscillations \citep[see, e.g.,][]{Wu_2003,Liu_Munoz_Lai_2015}.

In sum, we expect that physical collisions will be avoided if the binary enters the diffusive regime. In the remainder of this work, we focus on how frequently this occurs in galactic centers. 

\section{Reducing inner-orbit pericenters}\label{sec:dynamics_2}

\begin{figure}
    \centering
    \includegraphics[width=\linewidth]{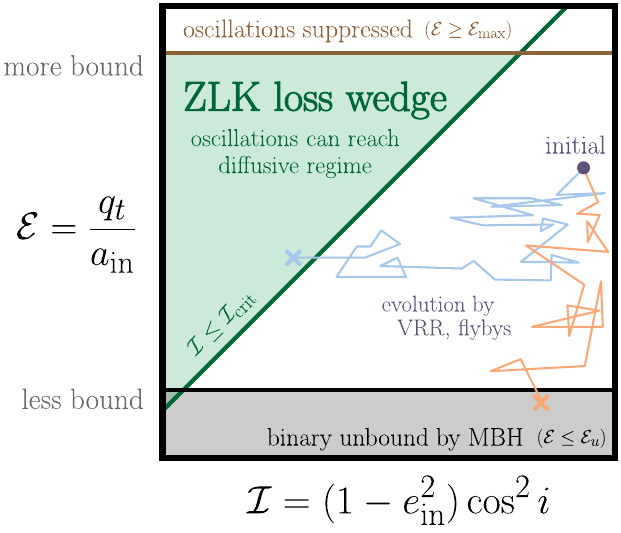}
    \caption{$\iom$ determines the maximum eccentricity a binary can reach during ZLK oscillations, while $\ene$ encodes the inner-orbit energy. 
    If $\iom$ is small enough for a given $\ene$, then we say the system is in the ``ZLK loss wedge'' (condition~\ref{eq:condition}; green region). Here, oscillations may bring the binary to the diffusive regime ($q_{\rm in} \leq q_t$; see eq.~\ref{eq:qt}), where it should tidally contract (Section~\ref{sec:fates}).
    Evolution in in $\ie$, driven by VRR and flybys, brings binaries into the loss wedge (e.g., blue path in cartoon); however, flybys may first widen the inner orbit to the point of becoming unbound (e.g., orange path; see eq.~\ref{eq:unbind}).}
    \label{fig:ie_space}
\end{figure}
\subsection{The ZLK loss wedge}\label{sec:ZLK_loss}

What fraction of systems will actually reach $q_{\rm in} \leq q_t$ through ZLK oscillations?
The minimal pericenter separation attined during an oscillation primarily depends on two parameters of the binary--MBH system.

First is the dimensionless ``ZLK constant''
\begin{equation}\label{eq:iom}
    \iom \equiv \frac{(J_{\rm in}\cos i)^2}{Gm_ba_{\rm in}} 
    = (1-e_{\rm in}^2)\,\cos^2 i,
\end{equation}
with $J_{\rm in}\equiv \sqrt{Gm_ba_{\rm in}(1-e_{\rm in}^2)}$. This value is conserved under quadrupole-order ZLK oscillations (and under octopole-order oscillations when $e_{\rm out} = 0$). 

Second is the dimensionless energy
\begin{equation}\label{eq:ene}
    \ene \equiv -E_{\rm in} \left(\frac{2q_t}{Gm_b}\right) = \frac{q_t}{a_{\rm in}}.
\end{equation}
This value is positive when the inner orbit has negative binding energy. It is also conserved under ZLK oscillations (so long as the system is hierarchical, i.e., $a_{\rm in} \ll q_{\out}[m_b/m_\bullet]^{1/3}$). 

In what part of $\ie$ space will oscillations bring a binary to the diffusive regime? 
When $\iom > 3/5$, a circular binary will remain circular. At smaller $\iom$, the minimum pericenter separation of an initially circular 
binary is given by \citep[e.g.,][]{Lidov_1962,Kozai_1962}
\begin{equation} \label{eq:qin_min}
    \qinmin = \frac{q_t}{\ene}\left[1-\left(1-\frac{5}{3}\iom\right)^{1/2}\right].
\end{equation}
For a non-circular orbit, the minimum also depends explicitly on initial eccentricity and argument of pericenter. Still, equation~(\ref{eq:qin_min}) serves as a useful approximation.

Short-range forces may suppress ZLK oscillations, as discussed at the end of Section~\ref{sec:fates}.  
These effects are most efficient at small $a_{\rm in}$, so they roughly present an $\ene_{\max}$ above which systems cannot reach the diffusive regime. GR-driven precession stabilizes a circular orbit against ZLK oscillations when \citep[Chapter 5.4.1,][]{Tremaine_book_2023}
\begin{equation}
 \frac{Gm_b^2 a_{\out}^3(1 - e_{\out}^2)^{3/2}}{c^2a_{\rm in}^4m_\bullet} > \frac{3}{4};
\end{equation}
this yields
\begin{align}\nonumber
    \ene_{\rm max} &\equiv q_t\left(\frac{3c^2m_\bullet}{4Gm_b^2a_{\out}^3(1 -e_{\out}^2)^{3/2}}\right)^{1/4} 
    \\
    &= 0.0223 
    \left(\frac{q_t}{0.013\,{\rm au}}\right)
    \left(\frac{m_b}{2\,M_\odot}\right)^{-1/2}
    \left(\frac{a_{\out}}{0.1\,{\rm pc}}\right)^{-3/4}.
\end{align}

Then for a binary to attain $\qinmin \leq q_t$ through ZLK oscillations alone, it must satisfy 
\begin{equation}\label{eq:condition}
    \iom \leq \iom_{\crit}(\ene) \simeq 
    \frac{6}{5}\ene \quad \text{and}\quad \ene \leq \ene_{\max}.
\end{equation}
We have dropped a term proportional to $\ene^2$ because binaries of interest have $\ene \ll 1$.

We can say that a binary satisfying condition~(\ref{eq:condition}) is in a ``ZLK loss 
wedge''\footnote{This name is based on the ``loss wedge'' \citep{Magorrian_1999}, present in loss cone problems with axisymmetric potentials. \citet{Chen_2009} introduced a ``Kozai wedge,'' which in our variables is $\iom \leq 2\ene$ when $\ene \ll 1$. For the qualitative work done here, this difference is negligible.}
(cf.~\citealp{Chen_2009}; see cartoon in Figure~\ref{fig:diagram}). 
If no other processes alter $\ie$ significantly over one oscillation timescale, a binary within the ZLK loss wedge will reach $q_{\rm in} \leq q_t$ and contract. 

Note that most binaries in the ZLK loss wedge have small-enough $\qinmin$ to collide --- a binary has $\qinmin \leq r_1+r_2$ if it further satisfies $\iom \leq \iom_{\crit}(\ene)\times(r_1+r_2)/q_t$, which comprises the majority of the wedge parameter space for many systems. As argued in Section~\ref{sec:dynamics}, such binaries should contract by diffusive tidal friction rather than collide.

Considering octopole-order terms in the ZLK Hamiltonian, systems with $e_{\rm out} > 0$ may ``flip'' in inclination across $\cos i = 0$ \citep[see][]{Naoz_2011,Naoz_2013,Katz_2011,Naoz_2016} after timescales longer than the oscillation period \citep[e.g.,][]{Antognini_2015,Weldon_2024}. Qualitatively, a system that flips takes a brief excursion into and back out of the loss wedge; during this excursion, the inner-orbit pericenter separation may or may not reach the criterion for diffusive tides. From \citet{Stephan_2016} --- assuming all of the mergers in the bottom panel of their Figure 8 come from such excursions --- we may estimate that octopole-order effects boost by a factor $\lesssim 1.5$ the fraction of systems reaching $q_{\rm in}\leq q_t$, relative to the fraction of systems born into the loss wedge (thin grey line of Figure~\ref{fig:fraction_vs_rout}). We expect flybys and VRR to provide a more substantial boost, as discussed in the following section. Hereafter we neglect octopole-order ZLK effects.

The ZLK loss wedge is a broadly applicable tool. Similar intuition is helpful for, e.g., compact-object binary mergers in hierarchical triples, or extreme mass ratio inspirals or (partial) tidal disruption events around MBH binaries. 

\subsection{Flybys and VRR}\label{sec:evolution}

So far we have considered the interplay between (quadrupole-order) ZLK oscillations and (diffusive) dynamical stellar tides. The parameters $\iom$ and $\ene$ are conserved for a binary outside of the ZLK loss wedge under such evolution. (A binary in the wedge will contract to $\ene \sim 1/2$.) 
A binary will undergo $e_{\rm in}$ oscillations with characteristic period $t_{\rm ZLK}$ (eq.~\ref{eq:t_osc}), unless it contracts.

Reality is complicated by other effects, especially in a galactic center. Flybys change a binary's inner-orbital elements 
near-instantaneously,\footnote{There is also a secular effect from stars passing at separations $\gtrsim v_pP_{\rm in}$ (where $v_p$ is the passing star's speed). Long-term evolution will be dominated by closer-passing flybys \citep[see, e.g., discussion in][]{Hamilton_Modak_2024}, where kicks are effectively impulsive; we focus on these close encounters.} 
yielding ``kicks'' in both $\iom$ and $\ene$. This will tend to reduce $\ene$ of a soft binary \citep[increase $a_{\rm in}$;][]{Heggie_1975}. As a characteristic flyby-evolution timescale, we use the average time for $\ene$ to change by order itself --- the ``evaporation'' timescale, $t_{\rm evap}$ \citep[e.g.,][]{Alexander_2014}. A binary is only formally unbound when $\ene \leq 0$; however, a wide orbit may not be stable against the tidal potential of the MBH, and may become disrupted \citep[e.g.,][]{Hills_1988,Yu_2003,Hopman_2009,Grishin_2017}. To account for such disruption, we approximate that any binary with
\begin{equation}\label{eq:unbind}
    \ene \leq \ene_u \equiv f\left(\frac{q_t}{q_{\rm out}}\right)\left(\frac{m_\bullet}{m_b}\right)^{1/3}
\end{equation}
will become unbound. We set the factor $f=2(3)^{1/3}$,
making this consistent with equation~(\ref{eq:a_wide}).

VRR continuously reorients the outer-orbit normal vector, changing the inclination $i$ and thereby $\iom$. 
This effect arises from a torque on the outer orbit induced by the non-smooth component of the cluster potential \citep{Rauch_1996}; this torque remains coherent, and sufficiently ``reshuffles'' the outer orbits, on a timescale $t_{\rm VRR}$ \citep[e.g.,][]{Rauch_1996,Kocsis_2015,Alexander_2017}. For times $< t_{\rm VRR}$, the change in $\iom$ is monotonic. For times $\gg t_{\rm VRR}$, the smooth evolution of $\iom$ traces out a path reminiscent of a random walk, with changes in direction and speed roughly every $t_{\rm VRR}$.

We neglect effects that alter the scalar angular momentum or energy of the outer orbit, such as scalar resonant relaxation \citep{Rauch_1996} and two-body relaxation \citep[e.g.,][]{Hopman_2009}. These effects act on timescales that are longer than the evaporation timescales for all binaries of interest (and often longer than a Hubble time; cf.~\citealp{Marklund_2025}).

\subsection{Combined dynamics: Analytical expectations}\label{sec:expectations}

\begin{figure}
    \centering
    \includegraphics[width=\linewidth]{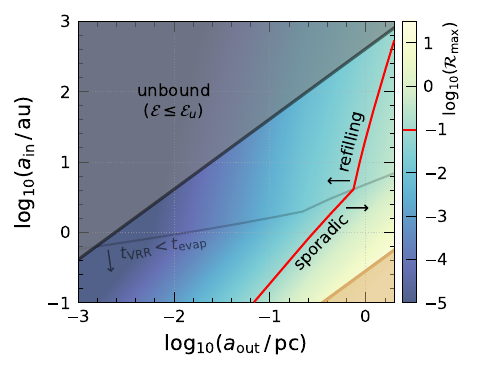}
    \caption{
    Binaries closer to the MBH tend to be solidly within the ``refilling'' regime, where contraction is more common (see Section~\ref{sec:regimes}). 
    This colormap shows $\rmax$, the ratio between the ZLK oscillation period and the faster of the evaporation and VRR timescales. The red contour denotes $\rmax = 0.1$, a rough boundary between the refilling and sporadic regimes. Below the grey contour, VRR acts faster than flybys. 
    In the grey region, binaries become unbound by the tidal potential of the MBH (i.e., they satisfy condition~\ref{eq:unbind}); in the gold region binaries are ``hard'' and will not evaporate. (All values are calculated for $m_b=2\,M_\odot$. Galactic Center properties are given in Appendix~\ref{sec:setup_details}.)}
    \label{fig:regimes}
\end{figure}

\subsubsection{Evolution in $\ie$}\label{sec:evo_in_ie}

We have established two important boundaries in $\ie$ space: the ZLK loss wedge (eq.~\ref{eq:condition}) and the unbinding energy (eq.~\ref{eq:unbind}). The probability of entering the loss wedge before becoming unbound --- and therefore the probability of contracting --- depends on the relative rate of evolution in $\iom$ and $\ene$. 

For a given binary, the characteristic timescale for evolution in $\ene$ is always $t_{\rm evap}$, as this evolution is driven by flybys alone. However, the timescale for evolution in $\iom$ is the minimum of $t_{\rm VRR}$ and $t_{\rm evap}$. Then if VRR acts more quickly than flybys ($t_{\rm VRR} \ll t_{\rm evap}$), a binary will be able to explore much of $\iom$ before evolving in $\ene$. Conversely, if flybys are faster ($t_{\rm VRR} \gg t_{\rm evap}$) \emph{or} if the two processes act at a similar rate, a binary will evolve in $\iom$ and $\ene$ on comparable timescales.

In the limit $t_{\rm VRR}/t_{\rm evap} \rightarrow 0$, all systems beginning with $\ene < \ene_{\max}$ should reach the ZLK loss wedge before becoming unbound. In the limit $t_{\rm VRR}/t_{\rm evap} \rightarrow \infty$, this fraction should go to $\sim 0.5$, as flybys induce an approximately unbiased random walk in $\ie$. While these arguments neglect higher-order effects --- like the importance of a system's initial position in $\ie$ --- they provide useful intuition. 

\subsubsection{Contraction in the loss wedge}\label{sec:regimes}

Once a system enters the ZLK loss wedge, it is not guaranteed to reach the diffusive regime ($q_{\rm in} \leq q_t$) and contract. To do so, it must remain in the wedge for $\sim$ a full ZLK cycle, so it can actually reach 
the minuimum pericenter of that cycle. Therefore, among systems that enter the wedge, contraction should be most common when ZLK oscillations are much faster than evaporation or VRR timescales.

Consider the timescale ratios $t_{\rm ZLK}/t_{\rm evap}$ and $t_{\rm ZLK}/t_{\rm VRR}$.
For a given binary, call the maximum of these ratios $\rmax$.
When $\rmax \ll 1$, oscillations are much faster than evolution in $\ie$. In Figure~\ref{fig:regimes}, we show the value of $\rmax$ over a range of inner- and outer-orbit semimajor axes.
There are two dynamical regimes of interest.

We say a system is in the ``refilling'' regime when $\rmax \ll 1$. This binary still effectively undergoes ZLK oscillations; however, successive oscillations follow trajectories determined by slightly different $\ie$. 

We say a binary is in the ``sporadic'' regime when $\rmax \gtrsim 1$. This binary's inner orbit no longer follows well-defined ZLK oscillations --- see examples in \citet{Hamers_2018} and \citet{Winter-Granic_2024} with large values of their respective
$\mathcal{R}$ parameters.\footnote{In \citet{Hamers_2018}, $\mathcal{R}=\rzv$. In \citet{Winter-Granic_2024}, $\mathcal{R}$ is a measure of the flyby-driven diffusion of $\pmb{j}_{\rm in} \equiv (1-e_{\rm in}^2)^{1/2}\pmb{\hat{J}}_{\rm in}$ during one eccentricity oscillation --- it therefore differs from $\rze$, though these definitions are equivalent at the extremes (i.e., the limit $0$ corresponds to only oscillations, while the limit $\infty$ corresponds to only flybys).}
Those works show that a sporadic-regime binary may still reach the diffusive regime, but it is no longer nearly certain that it will do so if it enters the loss wedge. 

We now discuss the expected fraction of contracted systems in these two regimes. Nearly every refilling-regime system that enters the loss wedge should contract; from Section~\ref{sec:evo_in_ie}, this should always be at least $\sim$ 1 in 2 systems.
In the sporadic regime, one can imagine that this fraction will be scaled by a reduction factor, proportional to the probability of remaining in the loss wedge for a full ZLK cycle. (Though the factor would also be proportional to the number of visits to the wedge, this number should never be greater than a few.) Assuming there is no strong preference for sporadic regime systems to reach the loss wedge, then, we expect contraction to be most common in the refilling regime.

\subsubsection{Radial dependence of contracted fraction}\label{sec:predictions}

Considering the arguments above, {the fraction of binaries that contract should be larger closer to the MBH.} Three main effects contribute to this radial trend.

First, recalling condition~(\ref{eq:condition}), the ZLK loss wedge spans more of $\ie$ space closer to the MBH. Binaries here must have larger $\ene$ (i.e., smaller $a_{\rm in}$), so $\iom_{\rm crit}$ is larger. Furthermore, short-range forces are less likely to suppress ZLK oscillations (i.e., $\ene_{\rm max}$ is larger) because the influence of the MBH is stronger. 

Second, binaries closer to the MBH are more likely to have $t_{\rm VRR} < t_{\rm evap}$, so they may more commonly enter the loss wedge regardless of its size (Section~\ref{sec:evo_in_ie}).

Third, crucially, binaries closer to the MBH are typically in the refilling regime. Therefore a larger fraction of the systems entering the loss wedge will actually reach the diffusive regime and contract (Section~\ref{sec:regimes}).

\section{Numerical experiment}\label{sec:setup}

We now carry out a simple numerical experiment on the dynamical evolution of binaries within the radius of influence of Sgr A*.
For each binary, we integrate secular equations of motion describing the evolution of the inner orbit (under quadrupole-order ZLK oscillations and GR precession) and the outer orbit (under VRR). We interrupt this secular evolution, at times determined by a Poisson process, to apply instantaneous perturbations to the inner orbit (mimicking the gravitational kicks from passing stars). We provide details of our exact setup in detail in Appendix~\ref{sec:setup_details}. 

We track each binary until one of three end states: 
\begin{enumerate}
    \item If $q_{\rm in} \leq q_t$, the binary \emph{contracts} (see eq.~\ref{eq:qt}).
    \item If $\ene \leq \ene_u$, the binary is \emph{unbound} (see eq.~\ref{eq:unbind}).
    \item If we reach the MS lifespan of the more-massive member of a binary, the binary is \emph{off the MS}.
\end{enumerate}

Importantly, in our simulations, \emph{we do not actually simulate the stellar-tide-driven evolution of the inner binary} --- we simply assume that binaries reaching $q_{\rm in} \leq q_t$ contract. 
So we are effectively modelling how dynamical processes in galactic centers cause binaries to reach small $q_{\rm in}$.

Once a binary has contracted, we assume that it does not continue to evolve dynamically. 

\subsection{Initial conditions}\label{sec:ics} 

We define $\rout$ as the outer-orbit-averaged distance between binary and MBH; this average separation is related to the outer-orbit semimajor axis and eccentricity by $\rout = a_{\out}{\left(1+e^2_{\out}/2\right)}$. We draw $\rout$ from a log-uniform distribution between $0.001$ pc and $2$ pc. We set $e_{\rm out} = 0$; we have verified that results with a thermal distribution are comparable. 

We draw $e_{\rm in} \sim \mathcal{U}(0,1)$ and $\cos i \sim \mathcal{U}(-1,1)$. We select the mass of the more-massive binary member, $m_1$, from an initial mass function (IMF) in which the number of stars with mass between $m$ and $m+dm$ is proportional to $m^{-\alpha}$, with 
$\alpha = 1.7$.\footnote{This is a top-heavy IMF relative to the solar neighborhood, as has been suggested for the Galactic Center \citep[e.g.,][]{Lu_2013}.}
We draw the mass ratio $m_2/m_1 \sim \mathcal{U}(0.1,1)$. We then draw $\log P_{\rm in}$ from a uniform distribution corresponding to semimajor axes $5(r_1 + r_2) < a_{\rm in} < q_t/\ene_u$. The lower limit ensures we do not start with binary members overflowing their Roche lobes, while the upper limit ensures that the binary is stable against the tidal force of the MBH (eq.~\ref{eq:a_wide}). 

We simulate $2\,874$ such binaries until they reach one of the end-state conditions listed above. Errorbars on each result show the variance of a binomial distribution with said result as the rate. 
\section{Results}\label{sec:results}

\begin{figure}
    \centering
    \includegraphics[width=\linewidth]{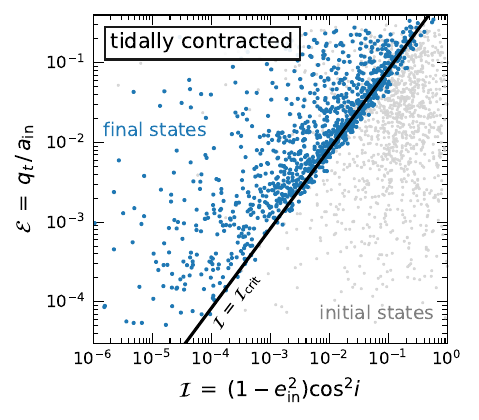}
    \caption{Binaries that contract (reach $q_{\rm in} \leq q_t$) mostly do so from within the ZLK loss wedge --- to the left of the black line (condition~\ref{eq:condition}). This figure only shows the initial (grey) and final (blue) states of systems that contracted.}
    \label{fig:loss_wedge_evidence}
\end{figure}

\subsection{Main results} \label{sec:depend_rout}

First, in Figure \ref{fig:loss_wedge_evidence}, we see that all binaries that contract do so from within --- or very close to --- the ZLK loss wedge. The imprecision of this boundary arises from our approximate definition of a ZLK oscillation's minimum pericenter separation (eq.~\ref{eq:qin_min}). No binary in our simulations contracted while having $\iom > 2\iom_{\rm crit}$.

\begin{figure}
    \centering
    \includegraphics[width=\linewidth]{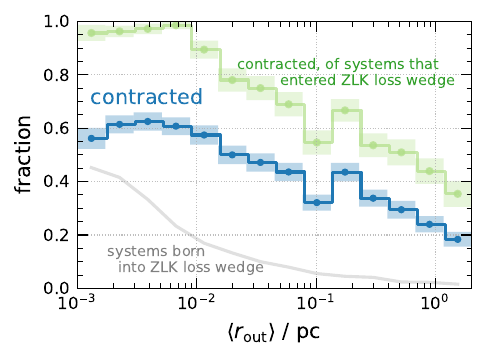}
    \caption{The fraction of binaries that tidally contract (blue) decreases with increasing average separation from the MBH, but is $\gtrsim$ 1 in 5 everywhere. There is a clear excess relative to the fraction born into the ZLK loss wedge (grey, shown for a larger sample of $50\,000$ binaries), though this excess is small at very small $\rout$.
    Close to the MBH, almost all systems entering the loss wedge end up contracting (green), as these systems are deep in the refilling regime.}
    \label{fig:fraction_vs_rout}
\end{figure}

The main results of this work are shown in Figure~\ref{fig:fraction_vs_rout}. We find that $\sim 3$ in 5 binaries at $\rout = 0.05$ pc undergo tidal contraction, and that this fraction declines by a factor of three by $\langle r_{\rm out}\rangle=1$ pc.

This radial dependence is consistent with our prediction in Section~\ref{sec:predictions}. We see that at least two of the three predicted contributing factors are borne out in our simulations: systems close to the MBH (1) are more likely to be born into the ZLK loss wedge (grey line), as the wedge spans more of the available $\ie$ space; and (2) are more likely to contract if they enter the loss wedge (green line), as they tend to be in the refilling regime (see Section~\ref{sec:regimes}). 

We had also discussed that systems closer to the MBH may be more likely to enter the wedge, based on our discussion in Section~\ref{sec:evo_in_ie}. This does not appear to be true --- in fact, we find the fraction of systems entering the loss wedge to be roughly consistent with 1 in 2 throughout the cluster. The cause of this is likely just that systems with $t_{\rm VRR} \ll t_{\rm evap}$ are very rare; our results are consistent with our discussion in Section~\ref{sec:evo_in_ie}. 

\subsection{Young nuclear population}

\begin{figure}[t]
    \centering
    \includegraphics[width=\linewidth]{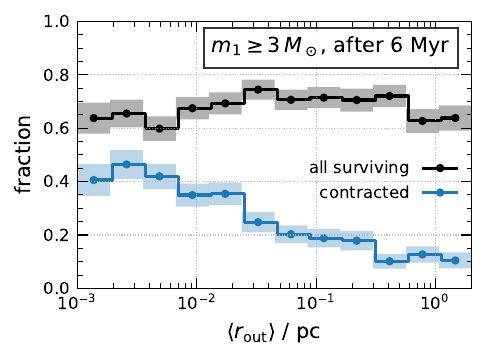}
    \caption{Fraction of binaries with $m_1 \geq 3\,M_\odot$ that are still bound after $6$ Myr --- roughly the age of the young stars in the Galactic Center. The total surviving binary fraction (black) is roughly constant across $\rout$, though the binaries in the inner region of the cluster are more likely to have contracted (blue). Further out, most surviving binaries retain roughly their original $a_{\rm in}$.}
    \label{fig:ync_observed}
\end{figure}

There appears to be a cluster of young stars within $\approx 0.5$ pc of Sgr A*, estimated to be $\approx 3$--6 Myr old \citep{Lu_2013}. Assuming this population contained binaries when it first formed, what fraction would we expect to still exist, contracted or otherwise?

In Figure~\ref{fig:ync_observed}, we show the fraction of binaries with massive primaries, $m_1 \geq 3\,M_\odot$, that remain bound after $6$ Myr of our fiducial simulation (black points). We call this the ``surviving fraction.'' We focus on massive (O-/B-type) primaries because they may be observable \citep[see, e.g.,][]{Gautam_2024}. The surviving fraction of these massive binaries is roughly 2 in 3, \emph{with little variation across} $\rout$. 
At smaller $\rout$, the surviving population is dominated by shrunken binaries, while at larger $\rout$, there are very few shrunken binaries because dynamical evolution is considerably slower. 

The surviving fraction's lack of dependence on $\rout$ is notable.
Unbinding becomes more prevalent than contraction at larger $\rout$; however, at any given time, the fraction of systems that have undergone neither process grows with $\rout$ (see, e.g., Figure 8 of \citealp{Stephan_2016}).
These two effects appear to cancel each other, yielding a surviving fraction that is roughly constant in $\rout$.

\subsection{Semimajor axis distribution}\label{sec:semimajor_axes}

\begin{figure}
    \centering
    \includegraphics[width=\linewidth]{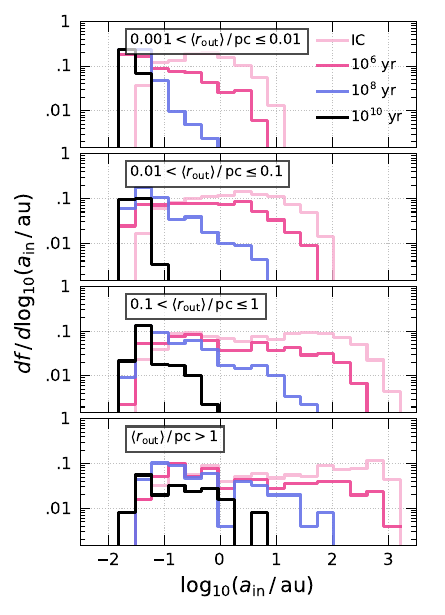}
    \caption{Distribution of inner-orbit semimajor axes in different bins of $\rout$. At later times, the distribution is dominated by contracted binaries, which accumulate at $a_{\rm in} \lesssim 0.1$ au. Different colored lines show different epochs, with light pink lines showing initial conditions. 
    Each histogram is normalized by the number of binaries initially in that bin of $\rout$, such that the ``IC'' histograms sum to 1 while all histograms at later times sum to the fraction of systems in that bin that are not unbound.}
    \label{fig:smas}
\end{figure}

The distribution of inner-orbit semimajor axes $a_{\rm in}$ varies with $\rout$ and with time. In Figure~\ref{fig:smas}, we show the $a_{\rm in}$ distribution, in different radial bins, at several epochs. For this, we prescribe contracted binaries a new $a_{\rm in} = 2q_t$. 

Over time, through contraction, the fraction of systems with $a_{\rm in} < 0.1$ au grows from negligible to dominant. The widest initial orbits are removed through both contraction and unbinding. Note that for $a_{\rm out} \lesssim 0.1$ pc, most contracted binaries are still dynamically soft; therefore they may evaporate on timescales ranging from 10 Myr to 10 Gyr post-contraction. Collisions with field stars may also destroy these binaries \citep[e.g.,][]{Fregeau_2004,Rose_2020}. Figure~\ref{fig:smas} does not account for the destruction of contracted systems, as we do not model these processes, but the main result of the figure would still hold.

\subsection{Time before contraction}\label{sec:time_to_shrink}

\begin{figure}
    \centering
    \includegraphics[width=\linewidth]{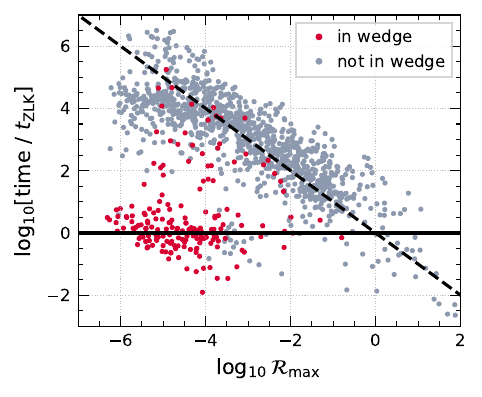}
    \caption{Binaries in the ZLK loss wedge tend to contract during their first ZLK cycle; binaries outside of it must dynamically evolve on $\ie$ first, so they tend to contract at times $\sim \min\left(t_{\rm evap},\,t_{\rm VRR}\right)$ (dashed line). Here, we plot the time it takes for each contracted binary to reach $q_{\rm in} \leq q_t$; each point is colored by whether or not the binary begins in the ZLK loss wedge (condition~\ref{eq:condition}).} 
    \label{fig:shrink_time}
\end{figure}

For each binary that reaches $q_{\rm in} \leq q_t$, we show the time it took to do so in Figure~\ref{fig:shrink_time}. (Once $q_{\rm in} \leq q_t$, the tide-driven contraction of $a_{\rm in}$ is rapid relative to $t_{\rm ZLK}$.) If the binary is in the wedge to start, it typically contracts after $\sim t_{\rm ZLK}$ . If it is out of the wedge, the time before contraction follows $\min\left(t_{\rm evap},\,t_{\rm VRR}\right)$ though there is a several-order-of-magnitude spread.

\subsection{Dependence on binary population parameters}\label{sec:other_depend}

The actual initial conditions of the Galactic Center binary population are highly uncertain. Fortunately, the results of a binary's dynamical evolution mainly depend on just four initial parameters. 

(1) The fraction of binaries that contract decreases with decreasing $\ene$ and 

(2) with increasing $\iom$. Binaries with smaller $\ene$ and larger $\iom$ are less likely to be born within the ZLK loss wedge, and they must ``travel'' further through $\ie$ space to reach the wedge, increasing the chance that they will be unbound or leave the MS first.

(3) The fraction of stars leaving the MS before contracting or being unbound increases with increasing primary star mass, $m_1$, due to the corresponding decrease in MS lifetimes and increase in evaporation timescale. The relative fraction of shrunken and unbound binaries does not depend strongly on $m_1$.

(4) As discussed in Section~\ref{sec:depend_rout}, the shrunken fraction decreases with increasing $\rout$. 

We find no dependence on mass ratio $m_2/m_1$, nor on $e_{\out}$ (though we note that higher-order/non-secular 3-body effects may increase the contracted fraction at $1-e_{\out} \lll 1$; e.g., \citealp{Stephan_2016,Stephan_2019,Mangipudi_2022}). Dependence on $e_{\rm in}$ and $i$ are subsumed by the dependence on $\iom$.

\section{Discussion}\label{sec:discussion}

\subsection{Takeaways}

The main purpose of this manuscript is not to provide rigorous numerical results on the evolution of the Galactic Center's stellar binary population. Rather, we aimed to clarify the interplay between the processes that sculpt such a population in any generic galaxy. 

To this end, we re-emphasize the ZLK loss wedge and associated refilling and sporadic regimes discussed in Section~\ref{sec:evolution}. When the ZLK oscillation period is much faster than the timescales for evaporation or VRR (refilling regime), tidal contraction while on the MS should be common (particularly when all timescales are shorter than MS lifetimes). In galaxies with more-massive MBHs, systems may be further into the refilling regime (see timescales in, e.g., \citealp{Hamers_2018}), so contraction may be more common. Specifically, while the contracted fraction will always decline with $\rout$, this decline will be shallower if the typical $\rmax$ at the radius of influence is smaller. 

The other purpose of this manuscript is to examine a model of binary evolution with starkly different approximations to those made in \citet{Stephan_2016,Stephan_2019}. As they (1) used an equilibrium-tides model and (2) did not model the dynamical effects of flybys and VRR, we find that their results (1) overestimated the fraction of collisions relative to contractions and (2) underestimated the fraction that undergo either of these two results. 

On point (1), the truth is likely between our results and theirs --- some fraction of the binaries that reach $q_{\rm in} \leq q_t$ may, indeed, collide rather than contracting. In Section~\ref{sec:fates}, we argued that this fraction should be small. Future work should continue to refine our treatment of stellar tides under ZLK oscillations (see also \citealp{Moe_2018,Marklund_2025};\footnote{Shortly before submission of this manuscript, we became aware of \citet{Marklund_2025}, which focuses on the dynamics of binaries with sub-Solar mass primaries near the hard/soft boundary at $\sim 0.1$ pc from Sgr A*. They consider the diffusive-tide regime, though their prescription does not allow for substantial contraction when a system reaches the regime through ZLK oscillations. We disagree with this implementation, so we expect that contraction is more common than their results suggest. That said, they perform 3-body integrations to account for flybys; for binaries near and below the hard/soft boundary, this is a substantial improvement over our impulse approximation (though we note that such binaries are rare in our cluster).}
see \citealp{Vick_2019} for an analogous problem with planets).

\subsection{Implications}

The predicted presence of a population of near-contact binaries has several implications; however, we cannot make strong, quantitative predictions, due to a serious lack of knowledge regarding the initial conditions of Galactic Center binaries. Nonetheless, we here mention four points of consideration.

(1) Near-contact binaries should necessarily interact (i.e., exchange mass, or undergo a common envelope phase) during post-MS evolution \citep[see, e.g.,][]{Sana_2012}. Relative to \citet{Stephan_2019}, we expect many fewer MS--MS mergers and many more post-MS mergers from radial expansion. Our results may then alter the expected distribution of Galactic Center X-ray binaries \citep[cf.][]{Hailey_2018,Mori_2021} or of ``primordial'' compact object binaries \citep[cf.][]{Tagawa_2020}, which may become LIGO/Virgo/KAGRA sources.

(2) Contracted systems may also become X-ray binaries through so-called ``exchange'' interactions, wherein a single, stellar-mass compact object encroaches on a dynamically hard binary and replaces one of the binary members \citep[e.g.,][]{Fregeau_2004,Ivanova_2008}. This should only occur at $a_{\rm out} \gtrsim 0.1$ pc, where contracted systems are hard. Exchange interactions have not been favored as a dominant producer of the Galactic Center X-ray binaries (see, e.g., Appendix C of \citealp{Generozov_2018}), largely due to the rareness of hard binaries in typical initial binary distributions. Our finding that 20--40 percent of binaries at $\rout > 0.1$ pc will contract makes this route more promising.

(3) Smaller-$a_{\rm in}$ binaries yield faster hypervelocity stars and shorter-period S-cluster stars \citep[e.g.,][]{Bromley_2006,Rossi_2014,Generozov_2020} if they undergo \citet{Hills_1988} mechanism disruptions. However, such disruptions typically come from $a_{\out}$ near or beyond the MBH radius of influence \citep[e.g.,][]{Lightman_1977,Yu_2003,Penoyre_2025}. We expect the contracted fraction to be very small at such distances. Therefore our results might \emph{not} alter expected hypervelocity/S-cluster star properties.

(4) The radial dependence of the surviving fraction at a given time (Fig.~\ref{fig:ync_observed}), combined with present-day observations of binaries in the young nuclear cluster, may provide insight into binary formation near Sgr A*. Our results suggest that binary survival at 6 Myr has little radial dependence. Furthermore, we should observe more of the surviving binaries at smaller $\rout$, as small-$a_{\rm in}$, contracted binaries should be easier to observe by eclipses or ellipsoidal variability \citep[see, e.g., selection functions in][]{Gautam_2024}. Meanwhile, observations suggest the present-day young binary fraction increases with increasing $\rout$ \citep{Chu_2023,Gautam_2024}. Together, these zeroth-order constraints require the \emph{initial} binary fraction to decrease sharply to smaller $\rout$. This conclusion is intuitive --- the increasing strength of the tidal field at smaller $\rout$ would increasingly suppress multiplicity --- and has been noted in simulations of star formation in MBH accretion disks (see, e.g., Section~4.5.7 of \citealp{Hopkins_2024}).

\subsection{Caveats and future work}\label{sec:caveats}

Several points of this study could be made more rigorous
(however, we emphasize again that our conclusions are most severely limited by a lack of knowledge of initial conditions, so added rigor is not currently valuable).
We list several of our major assumptions here. 

(1) We assumed that all binaries reaching the pericenter criterion for diffusive tidal evolution ($q_t$; eq.~\ref{eq:qt}) will contract immediately (cf.~Section~\ref{sec:fates}, Appendix~\ref{sec:tidal_onset}). 
(2) We neglected any effects that may arise when the double-averaged, quadrupole-order ZLK effect is not an appropriate approximation for the binary + MBH system dynamics \citep[cf.][]{Naoz_2016,Hamilton_2019a,Hamilton_2019b,Mangipudi_2022,Tremaine_2023}.
(3) We have assumed that stellar-mass perturbers have $m_\star=1\,M_\odot$ at all $\rout$ \citep[cf.][]{Panamarev_2019}.
(4) We have neglected direct collisions between binary members and background stars \citep[cf.][]{Rose_2020}.  
(5) We have assumed that the value of $q_t$ is constant over the MS lifetime of a star (cf.~Appendix~\ref{sec:tidal_onset}).
(6) We neglect the ``secular'' loss of $E_{\rm in}$ over many ZLK oscillations with $\qinmin > q_t$ \citep[cf.][]{Moe_2018}.
(7) We have limited our work to binaries on the MS.

Collectively, these assumptions (with the exception of [1], [3], and [4]) probably make our results conservative, i.e., the fraction of contracted binaries is even larger than we predict.

\section{Conclusion}\label{sec:conclusion}

We have studied the dynamical evolution of stellar binaries within the radius of influence of an MBH, using the inner few parsecs of the Galactic Center as a case study. We showed that of order 1 in 2 binaries here should contract to near-contact separations while still on the MS.

There is a ``ZLK loss wedge'' --- a region of the orbital parameter space where ZLK oscillations will bring a binary to sufficiently small pericenters to undergo this diffusive tidal contraction (Section~\ref{sec:ZLK_loss}). Only a small fraction of binaries are born within this wedge; however, passing stars perturb the binary and VRR changes the inclination between inner and outer orbits (Section~\ref{sec:evolution}). When these effects are slower than oscillations, they steadily ``refill'' the ZLK loss wedge, but when their timescales are comparable, they can also lead to ``sporadic'' instances of binary contraction (Section~\ref{sec:expectations}).

This confluence of effects means a large fraction of binaries reach pericenter separations of order a few stellar radii while still on the MS. We confirmed this through a numerical simulation incorporating ZLK oscillations, relativistic precession of the inner orbit, VRR, and flyby perturbations (Sections~\ref{sec:setup}--\ref{sec:results}). 

At such small pericenter separations, binaries should rapidly contract, as they reach the ``diffusive'' tidal regime (Section~\ref{sec:tidal_diff}). We provided a simple fitting formula for the onset of the diffusive-tide regime (eq.~\ref{eq:qt}). We then argued that systems reaching this regime would typically contract, rather than collide (Section~\ref{sec:fates}).

These considerations lead to our conclusion that a significant fraction of binaries within the radius of influence of an MBH should contract to separations of order a few stellar radii during their MS lifetimes. This fraction declines with separation from the MBH (Fig.~\ref{fig:fraction_vs_rout}), but is $\gtrsim$ 1 in 5 throughout the the Galactic Center. This lower limit may be higher in galaxies with more-massive MBHs. This may have implications for the post-MS evolution of nuclear cluster binaries, the properties of hypervelocity- and S-stars, and the inferred population of stellar binaries in the Galactic Center, which we touch on in Section~\ref{sec:discussion}. 

Lastly, while this work has focused on stellar binaries around an MBH, we emphasize that the ZLK loss wedge and the ``refilling'' and ``sporadic'' regimes of dynamical evolution introduced in Section~\ref{sec:dynamics_2} are entirely generalizable to other problems concerning tide-induced oscillations perturbed by external dynamics. Such systems are prevalent in astronomy, including bodies orbiting one member of a MBH--MBH binary, and field black hole triples. We emphasize that binaries in globular clusters should be subject to similar dynamics, both when they are in the presence of an IMBH and under the tidal potential of the cluster istelf.
\vspace{.2 cm}
\\
We thank Cristobal Petrovich, Smadar Naoz, Alexander Stephan, Dang Pham, and Fraser Evans for valuable conversations, and we thank the anonymous referee for their comments. We acknowledge support from NSERC grants RGPIN-2020-03885 and RGPIN-2024-05533.
\\
\textit{Software}: $\texttt{Python3}$ \citep{python}, $\texttt{scipy}$ \citep{scipy}, $\texttt{numpy}$ \citep{numpy}, $\texttt{pandas}$ \citep{pandas}, $\texttt{matplotlib}$ \citep{matplotlib}.
Proprietary code used in this work will be shared upon reasonable request to the corresponding author.

\bibliography{references}

\appendix

\section{Onset of diffusive tidal evolution}\label{sec:tidal_onset}

\subsection{Dynamical tides}

\citet{Press_1977} determined the energy deposited into a given normal mode $\alpha$ during a parabolic pericenter passage, $\Delta E_{\alpha}$. In response, the binary orbital energy changes by $\Delta E_{\rm in} = -\sum_{\alpha} \Delta E_{\alpha}$. This sum is dominated by the fundamental (f-) mode with quantum numbers $n=0$, $l=2$, $m=-2$. We call the f-mode frequency $\omega_f$. The fractional change to the inner-orbit energy from a pericenter passage is then \citep[e.g.,][]{Vick_2018,Wu_2018}
\begin{equation}\label{eq:delta_E}
    \left|\frac{\Delta E_{\rm in}}{E_{\rm in}}\right| \simeq \left|\frac{\Delta E_f}{E_{\rm in}}\right| = \frac{9W^2_{2-2}}{2\pi^2}\frac{q_m}{1+q_m}(1-e_{\rm in})^{-1}\left(\frac{q_{\rm in}}{r_1}\right)^{-2}Q_{02}^2 I_{2-2}^2\left(\frac{\omega_f}{\Omega_q}\right).
\end{equation}
We have defined the mass ratio $q_m \equiv m_2/m_1 \leq 1$ and the pericenter frequency $\Omega_q^2 \equiv Gm_b/q_{\rm in}$. The normalization parameter $W_{lm}$ (e.g., eq.~24 of \citealp{Press_1977}) is $W_{2-2}=(3\pi/10)^{1/2}$. To compute the dimensionless tidal overlap integral $Q_{02}$, 
we normalize our eigenfunctions such that $\int dV\rho[\pmb{{\xi}}_{\alpha}(\pmb{r})\cdot \pmb{{\xi}}_{\alpha}^*(\pmb{r})]=m_1$. 

For the orbit integral $I_{2-2}$, we use the expansion from Appendix~C of \citet{Lai_1997}. In this expansion, $I_{2-2} \propto z^{3/2}\exp(-2z/3)\left[1-(\pi/16z)^{1/2}\right]$, where $z \equiv \sqrt{2}\omega_f/\Omega_q \propto q_{\rm in}^{3/2}$ 
--- that is, at $q_{\rm in} >$ a few times $r_1$, the integral $I_{2-2}$ declines super-exponentially with increasing $q_{\rm in}$. The fractional energy change ($\propto I^2_{2-2}$) is then a \emph{very} steep function of pericenter separation, as seen in Figure~\ref{fig:n_to_shrink}.

\subsection{Criterion for diffusive tidal evolution}

Past works have shown that diffusive tidal evolution may occur at pericenter separations where the energy delivered to an f-mode at rest (i.e., with no initial amplitude) results in $|\Delta P_{\rm in}| \gtrsim \omega_f^{-1}$ \citep{Vick_2018,Wu_2018}.
In our notation,
\begin{align}\label{eq:delta_Phat}
    \omega_f|\Delta P_{{\rm in}}| 
    &= \tilde{\omega}_f \frac{27W_{2-2}^2}{2\pi} \frac{q_m}{(1+q_m)^{3/2}} \left(\frac{a_{\rm in}}{r_1}\right)^{5/2}\left(\frac{q_{\rm in}}{r_1}\right)^{-3} Q^2_{02} I_{2-2}^2\left(\frac{\omega_f}{\Omega_q}\right),
\end{align}
with dimensionless $\tilde{\omega}_f \equiv \omega_f (r_1^3/Gm_1)^{1/2}$.

We use \texttt{MESA} to construct zero-age MS models of stars with masses ranging from $0.3\,M_\odot$ to $100\,M_\odot$. For each model, we use \texttt{gyre} to solve for $\omega_f$ and $\xi_{(r,h)}$, which allow us to compute $Q_{02}$ and $I_{2-2}$. 
We then evaluate equation~(\ref{eq:delta_Phat}) for a range of $q_{\rm in}/r_1$ and $a_{\rm in}$ (taking $q_m = 1$ for simplicity). 

In Figure~\ref{fig:qt_num}, we show the value of $q_{\rm in}$ at which $\omega_f |\Delta P_{\rm in}| = 1$ --- this is $q_t$, the criterion for diffusive evolution. 
We present an approximate fitting formula in equation~(\ref{eq:qt}).
The dependence on $m_1$ primarily comes from corresponding changes to primary radius $r_1$, but differences in stellar structure also contribute. The roughly logarithmic dependence on $a_{\rm in}$ comes from the roughly exponential dependence of $\omega_f|\Delta P_{\rm in}|$ on $q_{\rm in}$.

\section{Details of numerical model}\label{sec:setup_details}

\subsection{ZLK Oscillations}

The equations of motion of $\pmb{e}_{\rm in}$ and $\pmb{j}_{\rm in}$ are those induced by the tidal potential of an MBH, expanded to quadrupole order in the ratio $(a_{\rm in}/a_{\rm out})$, averaged over many inner and outer orbits \citep[e.g., Chapter 5.4 of][]{Tremaine_book_2023}. We add the first-order post-Newtonian precession term,
\begin{equation}\label{eq:gr_prec}
    \left(\frac{d\omega}{dt}\right)_{\rm GR} = \frac{3{(Gm_b)}^{3/2}}{a^{5/2}_{\rm in}{\left(1-e_{\rm in}^2\right)}^{3/2}c^2}.
\end{equation}

For simplicity, we neglect octopole-order terms in the expansion of $(a_{\rm in}/a_{\rm out})$ \citep[see, e.g.,][]{Naoz_2013,Naoz_2014,Naoz_2016}, as well as the additional terms in Brown's Hamiltonian \citep{Brown_1936a,Brown_1936b,Brown_1936c} that may be relevant in cases such as ours, where $m_\bullet \gg m_b$ \citep[see][]{Tremaine_2023}. Including these terms --- or using a Hamiltonian only averaged over the inner-orbit period --- may serve to increase the fraction of systems reaching very small $q_{\rm in}$ \citep[see also][]{Mangipudi_2022}. 
We tested the influence of a spherical Hernquist potential (e.g., \citealp{Winter-Granic_2024}; see \citealp{Hamilton_2019a,Hamilton_2019b} for generic cluster tide equations), but we found that the effects of this deviation from quadrupole-order ZLK evolution are negligible. 

\subsection{Vector resonant relaxation}

Under the influence of VRR, the orientation of the outer angular momentum vector (${\pmb{j}}_{\out}$) evolves stochastically. 
For our toy model, we prescribe $d\pmb{j}_{\out} / dt = \pmb{\eta} \times \pmb{\hat{\jmath}}_{\out}$, with $\pmb{\eta}$ a time-varying, 3$D$ Gaussian noise vector \citep[cf.][]{Hamers_2018,Fouvry_2019}.
At $t = nt_{\rm VRR}$ for some integer $n$, we set the $i$th component of the vector ${\eta}_i = T_na_{ni}$. The torque strength $T_n \sim \mathcal{N}(0,T_{\rm rms})$, drawn at each $n$, where we say
\begin{equation}
    T^2_{\rm rms} = 0.18\pi g_\gamma\frac{N_\star}{P_{\out}^2}\left(\frac{m_\star}{m_\bullet}\right)^2
\end{equation} 
is the variance of the VRR torque (cf.~eqs.~35, 38, and 48 of \citealp{Fouvry_2019}; $g_\gamma$ depends on the cluster density power law parameter and is given by their eq.~47). We also draw the amplitudes $a_{ni} \sim \mathcal{N}(0,1)$ at each $n$. We interpolate between the $\pmb{\eta}$ vectors drawn at $t=(n-1)t_{\rm VRR}$ and at $nt_{\rm VRR}$ using logistic functions, with smoothing parameter $k = 50$. This model yields statistically accurate short- and long-term evolution of the outer orbit orientation.

\subsection{Encounters}

We model interactions between a binary and other cluster members using the impulse approximation. For a perturber of mass $m_p$, with speed $v_p$, that comes within a distance $b$ of the binary barycenter, this requires (1) that the timescale of the interaction $b/v_p \ll P_{\rm in}$ and (2) that the path of the perturber is negligibly altered by the interaction (typically, $v_p^2 \gg G[m_b+m_p]/b$). With the large velocity dispersion of galactic center environments, these conditions are typically satisfied out to $\sim 10 a_{\rm in}$.

Under the impulse approximation, the positions of the binary members ($\pmb{r}_i$) do not change. The velocity vector of each receives a kick
\begin{equation}
    \Delta\pmb{v}_i = \frac{2Gm_p}{v_p}\frac{\pmb{\hat{b}}_i}{b_i},
\end{equation}
where $\pmb{b}_i$ is the minimum-separation vector between the perturber and the $i$th binary member.

We consider perturbations from stars coming within $d=10a_{\rm in}$ of the inner-orbit barycenter. The time between interactions is drawn from an exponential distribution with scale parameter $t_{\rm enc} = (n\pi d^2\sigma_v)^{-1}$, which yields Poisson-distributed encounter statistics. We take all perturbers to have $m_p = 1\,M_\odot$.
We choose the flight paths of perturbers such that they approach the binary isotropically \citep[see][]{Henon_1972}, with minimum distance to the inner-orbit barycenter following a probability density function $p(b)db \propto bdb$ for $b\leq d$. 

Choosing appropriate flight paths based on these requirements, we can determine the appropriate $\Delta \pmb{v}_i$ for an interaction. We then update the vectors $\pmb{j}_{\rm in}$ and $\pmb{e}_{\rm in}$ \citep[see also][]{Collins_2008} and return to secular evolution until the next encounter arrival time.

\subsection{Galactic Center properties}
The MBH and cluster parameters determine the efficiency of ZLK, VRR, and flybys. We set the MBH mass to $m_\bullet = 4\times10^6\;M_\odot$ \citep{Gravity_2023}. To determine the local velocity dispersion $\sigma_v$, the local mass density $\rho$, and the number of stars $N_\star$ interior to each binary, we evaluate equations~(\ref{eq:dense})--(\ref{eq:velo_disp}) at $r = \rout$.

The density profile is given by
\begin{equation}\label{eq:dense}
    \rho(r) = \rho_0 {\left(\frac{r}{r_0}\right)}^{-\gamma_i},
\end{equation}
with $\rho_0 = 2.8\times10^6\,M_\odot\;{\rm pc}^{-3}$ and $r_0 = 0.22\,{\rm pc}$; the power law slope is $\gamma_1 = 1.2$ for $r < r_0$ and $\gamma_2 = 1.75$ for $r \geq r_0$ \citep{Lockmann_2009}. The number of stars internal to $r$ is then given by computing 
\begin{equation}\label{eq:nstar}
    N_\star(r) = \frac{4\pi}{m_\star}\int_0^r dr'\rho(r')r'^2.
\end{equation}

\citet{Kocsis_2011} find the one-dimensional velocity dispersion by solving the requisite Jeans equation (\citealp{BT_2008}, eq.\ 4.216), assuming the velocity dispersion tensor of the nuclear cluster is isotropic. Their solution is approximated (to errors smaller than the observational uncertainties) by
\begin{equation}\label{eq:velo_disp}
    \sigma_v(r) = 250\,{\rm km\; s}^{-1} {\left(\frac{r}{0.1\,{\rm pc}}\right)}^{-\frac{1}{2}}.
\end{equation}

\end{document}